\documentstyle[prb,twocolumn,aps]{revtex}
\tighten
\begin{document}
\draft
\title{Chaos in resonant-tunneling superlattices}
\author{O. M. Bulashenko\cite{byline} and L. L. Bonilla}
\address{
Universidad Carlos III de Madrid, Escuela Polit\'{e}cnica Superior,
Butarque 15, 28911 Legan\'{e}s, Spain }

\twocolumn[
\date{\today}
\maketitle
\widetext \vspace*{-1.0truecm}

\begin{abstract} \begin{center} \parbox{14cm}
{Spatio-temporal chaos is predicted to occur in n-doped semiconductor
superlattices with sequential resonant tunneling as their main charge transport
mechanism. Under dc voltage bias, undamped time-dependent oscillations of the
current (due to the motion and recycling of electric field domain walls) have
been observed in recent experiments.
Chaos is the result of forcing this natural oscillation by means of an
appropriate external microwave signal.}
\end{center} \end{abstract}


]\narrowtext

Nonlinear oscillations and chaos have been predicted in systems where a
predominantly quantum dynamics is corrected by mean-field nonlinear terms due
to
collective interactions (Hartree) \cite{capasso92} or to interactions with
classical subsystems having a widely different time scale. \cite{BG}
These phenomena are different from quantum chaos, \cite{haake} i.e., the
behavior of quantum systems whose classical counterpart is chaotic.
So far chaotic oscillations have been predicted for systems with few degrees of
freedom and experimental evidence is scarce.
In this paper we predict chaotic behavior with loss of spatial coherence in a
system with many degrees of freedom for which the main transport mechanism is
resonant tunneling: a weakly-coupled multiquantum well superlattice (SL).
As far as we know these are the first results on chaotic behavior in SLs.
In contrast to unbiased triple-well heterostructure, \cite{capasso92} our SL is
subject to external $dc$+$ac$ bias, as it was the case with the two-level
system
of Ref.\ 2.

Very recently time-dependent oscillations of the current on GaAs/AlAs SLs
subject to {\em dc} voltage bias have been found. \cite{Kwok}
The oscillations are damped for undoped photoexcited SLs and undamped for doped
SLs without photoexcitation. \cite{Kastrup}
For large values of the photoexcitation or the doping, there is stable
formation
of stationary electric field domains leading to the well-known oscillatory
\mbox{$I$-$V$} characteristic. \cite{Grahn91,lamiller,Prengel}
According to a discrete drift model, \cite{Bonilla94} the current oscillations
are caused by the creation, motion and recycling of domain walls separating two
electric field domains. \cite{Bonilla-ICPF}
This situation is reminiscent of that found in bulk semiconductor devices with
negative differential resistance (NDR), where {\em dc} voltage bias gives rise
to high-field domain dynamics and the well-known Gunn oscillations.
\cite{shaw,higuera92}
A significant difference is that the space charge waves are dipoles in the Gunn
oscillations and charge monopoles in the SL current oscillations.
Another is that the Gunn waves are generated close to the injecting contact
whereas the domain walls appear clearly inside the SL. \cite{Bonilla-ICPF}

Having found a system with a natural oscillation due to traveling-wave motion,
it is natural to ask whether harmonic forcing would lead to chaos with spatial
structure. The answer is affirmative.
This is also the case for the periodically driven Gunn diode studied by
Mosekilde {\it et al}. \cite{mosekilde}
Experimental studies have been carried out by Kahn {\it et al.} \cite{kahn92}
on
ultrapure p-Ge, where the NDR is caused by negative differential impurity
impact
ionization, \cite{BT} and the transition to a chaotic attractor with loss of
spatial coherence has been observed.

We consider a set of weakly interacting quantum wells (QWs) characterized by
average values of the electric field ${\cal E}_i(t)$, and the electron density
$n_i(t)$, with $i=1,\dots,N$ denoting the QW index.
This mean-field-like approach is often justified because the relevant time
scale
for the oscillations ($\sim 0.1~\mu$s) is much larger than those for the
tunneling process between adjacent QWs ($\sim 1$~ns) and the relaxation from
excited levels to the ground state within each QW ($\sim 1$~ps).
\cite{Bonilla94} The one-dimensional equations governing the dynamics of
the system are the Poisson equation averaged over one SL period $l$, Amp\`ere's
equation for the balance of current density and the voltage bias condition:
\begin{eqnarray} \label{pois}
\frac{1}{l} \; ({\cal E}_i - {\cal E}_{i-1}) = \frac{e}{\epsilon} \;
(n_i - N_D) \\
\label{amp}
\epsilon \; \frac{d {\cal E}_i}{d t} + e n_{i} v({\cal E}_i) = J \\
\label{bias}
l \sum_{i=1}^{N} {\cal E}_i = V(t).
\end{eqnarray}
Here $\epsilon$, $e$, and $N_D$ are the average permittivity, the electron
charge, and the average doping density, respectively. \cite{remark}
The total current density $J(t)$ is the sum of the displacement current and
the electron flux due to sequential resonant tunneling $e n_i v({\cal E}_i)$.
The effective electron velocity $v({\cal E})$ exhibits maxima at the resonant
fields for which the adjacent levels of neighboring QWs are aligned,
\cite{remark2} as shown in the inset of Fig.\ \ref{poincare}.
The voltage $V(t)$ in (\ref{bias}) is the sum of a $dc$ voltage $V_b$ and an
$ac$ microwave signal of relative amplitude $A$ and driving frequency $f_d$:
$V(t) = V_b \{1+A \sin (2 \pi f_d t) \}$.
The boundary condition $\epsilon ({\cal E}_1 - {\cal E}_0)/(el) = n_1 - N_D =
\delta$ allows for a small negative charge accumulation in the first well,
which
is taken to be $\delta\sim 10^{-3}\times N_{D}$.
The physical origin of $\delta$ is that the n-doped SL is typically sandwiched
between two n-doped layers with an excess of electrons, thereby forming a
n$^+$-n-n$^+$ diode. \cite{Kastrup}
Then some charge will be transferred from the contact to the first QW creating
a
small dipole field that will cancel the electron flow caused by the different
concentration of electrons at each side of the first barrier.

To study our equations, it is convenient to render them dimensionless, using
the
characteristic physical quantities.
As the unit of the electric field $E={\cal E}/{\cal E}_{1-2}$ we adopt the
typical electric field strength needed to align the first and the second
electron subbands of the nearest QWs ${\cal E}_{1-2}$.
The dimensionless velocity ${\rm v}(E)$ is obtained normalizing $v({\cal E})$
by its value at ${\cal E}={\cal E}_{1-2}$, where it has a local maximum due to
resonance in tunneling (see inset of Fig.\ \ref{poincare}).
The others dimensionless quantities are defined as follows:
the donor concentration $\nu = e l N_D/ \epsilon {\cal E}_{1-2}$;
the time $\tau = t/t_{\rm tun}$ where $t_{\rm tun}=l/v({\cal E}_{1-2})$ is the
characteristic tunneling time; the $dc$ bias ${\cal V} = V_b / {\cal E}_{1-2} l
N$; the $ac$ bias amplitude $a=A {\cal V}$;
the driving frequency $\omega = 2 \pi f_d t_{\rm tun}$.
By time-differentiating (\ref{bias}) and using Amp\`{e}re's law (\ref{amp}),
one
can express the current density as
\begin{equation} \label{curr}
J(t) = \frac{\epsilon}{l N} \frac{d V}{d t}
+ \frac{e}{N} \sum_{j=1}^{N} n_j v({\cal E}_j),
\end{equation}
and after its substitution into
(\ref{amp}) we obtain a system of $N$ equations for the electric field
profiles:
\begin{eqnarray} \label{map}
\frac{d E_i}{d \tau} = && \frac{1}{N}\sum_{j=1}^{N} {\rm v}(E_j) \;
[ E_j -E_{j-1} + \nu ] \nonumber \\ &&{}
- {\rm v}(E_i) \; [ E_i -E_{i-1} + \nu ] + a \omega \cos (\omega \tau),
\end{eqnarray}
which we solved numerically by the fourth-order Runge-Kutta method with the
boundary condition $E_0 = E_1 - \nu\delta$ and initial conditions $E_i(0) =
{\cal V},\, \forall i$.
As an example, we consider a GaAs/AlAs SL at $T$=5 K with $N$=40, $l$=13 nm,
${\cal E}_{1-2}\approx$10$^5$ V/cm, $N_D \approx$1.15$\times 10^{17} {\rm
cm}^{-3}$, for which undamped time-dependent oscillations of the current were
first observed. \cite{Kastrup} One get $\nu\approx 0.1$, $t_{\rm tun}\approx$
2.7 ns, and we take ${\cal V}$=1.2 (corresponding to $V_b\approx$7.8 V).
The main features of our numerical results are as follows.

For the $dc$ case ($a$=0) undamped time-periodic current oscillations of
frequency $f_0\approx 9$ MHz, (in excellent agreement with the observed value
\cite{Kastrup}) set in after a transient period.
The electric field and charge profiles corresponding to one period of the
current oscillation are similar to those found in the undoped case
(cf.\ Figure 9 in Ref.\ 9).
During each period a domain wall (charge monopole) is formed inside the SL.
It then moves towards the corresponding contact.
Depending on the applied voltage, it may or may not reach the end of the SL
before it dissolves and a new monopole is formed starting a new period of the
oscillation. Since in a 40-period SL the wall moves only a few QWs, it may be
hard to distinguish this monopole recycling from a true spatial oscillation of
a
single monopole. Simulations of longer SLs ($N\geq$100), however, clearly show
monopole recycling with two monopoles coexisting during some part of one
current
oscillation period. The frequency of the oscillation is mainly determined by
the number of QWs the monopole moves across (which increases with $N$) and by
the average drift velocity. \cite{Bonilla-ICPF}

Now consider the $ac$ case. We start with a uniform initial field profile and
solve the equations for $dc$ bias. After a short transient, the self-sustained
oscillations set in and we switch on the $ac$ part of the bias.
Our main result is that the competition between the natural oscillation due to
monopole dynamics and the forcing gives rise to narrow windows of
spatio-temporal chaos for appropriate values of ${\cal V}$, $a$ and $\omega$,
and that the size and richness of these windows increases with $N$. Let us fix
the ratio between the natural frequency $f_0$ and the driving frequency $f_d$
at
the golden mean $(\sqrt{5}-1)/2$.
To detect and visualize the chaotic regions in parameter space, we need to
define a Poincar\'e mapping. The current is a good measure of the amplitude
(norm) of the solutions which is illustrated by the use of current vs voltage
characteristics as bifurcation diagrams. \cite{higuera92}
Denoting the period of the $ac$ bias $T_d=2 \pi/\omega$, we adopt as Poincar\'e
mapping (for each value of $a$) the current at times $\tau_m=m T_d$, $m = 0, 1,
\ldots$ (after waiting enough time for the transients to have decayed).
The result is the bifurcation diagram in Fig.\ref{poincare}.
Notice the period-doubling sequences that point to the existence of chaos
near their accumulation points. There we have computed the largest
Lyapunov exponent and found it positive, which confirms chaos within the
windows marked by arrows in Fig.\ref{poincare}. The chaotic
regions are interspersed with locking to periodic and quasiperiodic regimes.
Quasiperiodic routes to chaos have been found at the first and last windows
marked with arrows in Fig.\ref{poincare}. Notice that the period-2 orbits span
the widest parameter region from the
narrow chaotic band around $a \approx 0.01$ up to the next chaotic region at $a
\approx 0.09$.  For $a \ge 0.145$ the solution is attracted to the period-1
orbit with the driving frequency $f_d$.

More insight into the transition between chaotic and non-chaotic regions of
the bifurcation diagram can be obtained by sweeping-down through it, as
follows.
We set $a_0$=0.16 (the leftmost value in Fig.\ \ref{poincare}, where there is
period-1 locking) and take an initial field profile $\{E_i\}={\cal V}$.
We now solve the problem (\ref{map}), compute $J_m = J(m T_d)$ and find out
when
the solution is periodic within a $10^{-5}$ accuracy. At that time, we stop the
simulation, depict all $J_m$ correspondent to the period, store the resulting
electric field profile, and use it as initial condition for another simulation
with $a = a_0 - \Delta a$, $\Delta a = 2\times 10^{-4}$. By repeating this
process, we arrive at chaotic or quasiperiodic regions of $a$, where the
integration is stopped at a much larger time $mT_d$, with $m$=2000, and only
the
last 1000 points are depicted, thereby eliminating transients.
Fig.\ref{poincare} is the result of such a sweep-down run.
The transition points between different regions in the bifurcation diagram
are found to be different when a sweep-up run is made, demonstrating thus
hysteresis.

To illustrate the spatially chaotic nature of the solutions we
pick arbitrarily two far-away QWs and depict the simultaneous values of
the electric field at them after each period of the driving force $T_d$.
The resultant attractors for several amplitudes $a$ are presented in Fig.\
\ref{ee}. The first example [Fig.\ \ref{ee}(a)] shows a chaotic attractor
with layered structure and variation in the density of points. The chaotic
attractor of Fig.\ \ref{ee}(b) has several separate branches almost
continuously filled.  Finally Fig.\ \ref{ee}(c) corresponds to a value of
$a$ on the quasiperiodic region. The closed loop with periodic pattern [see
Fig.\ \ref{ee}(d)] indicates quasiperiodicity: the orbit fills the attractor
(torus), never closing on itself. The points in the chaotic attractors cluster
with varying density on different regions, which means that they can be
characterized by their multifractal dimension $D_q$ (calculated by
the method of Ref.\ 18).
We have found that for the particular case of Fig.\ \ref{ee}(a), $D_q$
decreases
from $D_{- \infty}\approx$ 1.57 to $D_{+ \infty}\approx$ 0.72, going through
the
capacity dimension $D_0 \approx 1.18$.

Loss of spatial coherence in the chaotic regime is easily found in long SLs.
Density plots for the electron concentration (dark regions mean high electron
density) in a 200-period SL with natural frequency $\approx$ 1.9 MHz are shown
in Fig.\ \ref{st}. The density plots show the transition from periodic [Fig.\
\ref{st}(a)] to chaotic domain wall dynamics with loss of spatial coherence
[Fig.\ \ref{st}(b)]. Under $dc+ac$ driving with golden mean frequency ratio,
nucleation of monopole wavefronts occurs more frequently: in addition to
long-living waves travelling over almost the whole SL, there are short-living
waves. The two types of waves are distributed chaotically in space and
nucleated
both at the beginning and deep inside the SL. Detailed calculations show
coexistence of up to three monopoles connecting four electric field domains
during certain time intervals. We thus have that the loss of spatial coherence
is due to chaotic domain-wall dynamics, as seen in Fig.\ \ref{st}(b).

In conclusion, spatio-temporal chaos is expected to occur in weakly-coupled
semiconductor superlattices (with sequential resonant tunneling as the main
transport mechanism) under appropriate $dc$+$ac$ voltage bias. This prediction
should be experimentally testable in currently available n-doped GaAs/AlAs
samples forming n$^+$-n-n$^+$ diodes. \cite{Kastrup}

We thank J.\ M.\ Vega and C.\ Martel for valuable discussions. O.\ M.\ B.\
has been supported by the Ministerio de
Educaci\'on y Ciencia of Spain. This work has been supported by the DGICYT
grant PB92-0248, and by the EC Human Capital and Mobility Programme contract
ERBCHRXCT930413.

\begin{figure}
\caption{Bifurcation diagram of the current at times $m T_d$ (for sufficiently
large $m$) versus the driving-force amplitude, for a 40-period SL and the
golden-mean ratio between natural and driving frequencies. Windows of chaotic
solutions are marked by arrows.
Inset: Dimensionless velocity as a function of the electric field. The point
indicates the electric-field value corresponding to the dc bias ${\cal V}=1.2$
used in the calculations.
}\label{poincare}\end{figure}

\begin{figure} \caption{
Electric field values $E_5$ and $E_{36}$ (at two different QWs of
the 40-period SL) measured at times $m T_d$ for:
(a) $a$=0.088; (b) $a$=0.101; (c) $a$=0.14. (d) is a blow-up of the small
region inside the rectangle in (c). $10^4$ successive points have been used
to depict each attractor.
}\label{ee}\end{figure}

\begin{figure} \caption{
Density plots for the electron concentration in a 200-period SL for:
(a) pure $dc$ bias; and (b) $dc+ac$ bias with $a$=0.112.
$x$- and $y$-axes correspond to time (in units of 16 $t_{\rm tun}$) and QW
index respectively. Darker regions indicate high electron density and mark the
location of the domain boundary for each time.
}\label{st}\end{figure}


\begin{thebibliography}{99}
\bibitem{remark}
The doping density $N_D$ has been incorporated into the equations of Ref.~9,
and the hole density has been set equal to zero, as we do not include
photoexcitation due to laser illumination in this paper.
\bibitem{remark2}
In our calculations we use the curve $v(\cal E)$ obtained from the experimental
current-voltage characteristics of Ref.~5, as explained in Ref.~4.
\end{thebibliography}
\end{document}